\documentclass[12pt]{article}

\usepackage{latexsym,amsmath,amssymb,theorem,epsfig}

\usepackage{graphicx}
\usepackage{graphicx,subfigure}
\usepackage{epstopdf}
\usepackage{amssymb}
\usepackage{amsmath}
\usepackage{psfrag}
\usepackage{epsfig}
\usepackage{cancel}

 \allowdisplaybreaks[4]

\usepackage[all]{xy}

\DeclareFontFamily{OT1}{rsfs10}{}
\DeclareFontShape{OT1}{rsfs10}{m}{n}{ <-> rsfs10 }{}
\DeclareMathAlphabet{\mathscript}{OT1}{rsfs10}{m}{n}

\numberwithin{equation}{section}

\newcommand{\ns}{\normalsize}

\newcommand{\cH}{{\cal H}}

\newcommand{\vd}{\dot V}
\newcommand{\td}{\dot \theta}

\def\gsim{ \lower .75ex \hbox{$\sim$} \llap{\raise .27ex \hbox{$>$}} }
\def\lsim{ \lower .75ex \hbox{$\sim$} \llap{\raise .27ex \hbox{$<$}} }
\def\be{\begin{equation}}
\def\ee{\end{equation}}
\def\bea{\begin{eqnarray}}
\def\eea{\end{eqnarray}}

\usepackage{graphicx}

\newcommand{\ba}{\begin{array}}
\newcommand{\ea}{\end{array}}

\begin{document}

\begin{titlepage}
\title{
  \hfill{\ns }  \\[1em]
   {\LARGE Gauss-Codazzi Thermodynamics \\ on the Timelike Screen}
\\[1em] }
\author{Federico Piazza
     \\[0.5em]
   {\ns Perimeter Institute for Theoretical Physics, Waterloo, Ontario, N2L 2Y5, Canada}\\[0.3cm]}

\date{}

\maketitle

\begin{abstract}
It is a known result by Jacobson that the flux of energy-matter through a local Rindler horizon is related with the expansion of the null generators in a way that mirrors the first law of thermodynamics. We extend such a result to a timelike screen of observers with finite acceleration. Since timelike curves have more freedom than null geodesics, the construction is more involved than Jacobson's and few geometrical constraints need to be imposed: the observers' acceleration has to be constant in time and everywhere orthogonal to the screen. Moreover, at any given time, the extrinsic curvature of the screen has to be flat. The latter requirement can be weakened by asking that the extrinsic curvature, if present at the beginning, evolves in time like on a cone and just rescales proportionally to the expansion.
\end{abstract}

\thispagestyle{empty}

\end{titlepage}

\section{Introduction}

Various analogies between gravity and thermodynamics have been pointed out which are based on the black hole entropy-area law~\cite{bek,bardeen}. By applying Raychaudhuri equation to a congruence of initially non-expanding light-like curves, Jacobson showed~\cite{jac} that the lensing effect of any matter-energy flux follows the first law of thermodynamics,
\begin{equation} \label{first}
d E = T dS,
\end{equation}
once $T$ and $d E$ are interpreted as the Unruh temperature and matter-energy flux seen by an accelerated observer and $dS$ is the change of area spanned by the null curves living on its Rindler horizon. Eq.~\eqref{first} is obtained in the limit of infinite acceleration, in which the observer's worldline gets infinitely close to the Rindler horizon. Hence, effectively, Jacobson approach (see~\cite{lev,cai} for cosmological applications), deals with the dynamics of null surfaces. Such a point of view has been deepened by Padmanabhan in a series of works (see {\emph{e.g.}~\cite{pad}), where the thermodynamics of horizons has been studied in great detail and other analogies explored.

The main novelty of a recent paper by Verlinde~\cite{ver} (see also \cite{nic}seems that of using arbitrary stationary \emph{timelike surfaces}\footnote{This Lorentz-breaking choice is consonant with the description of the regions of space as quantum subsystems that has been suggested by the present author and collaborators in a number of papers~\cite{glimmers,sergio,review} and that is based, from the very beginning, on a non-covariant splitting into ``space+time".}  -- rather than horizons --  as the boundary of the system/region of space taken into consideration.
The choice of a timelike screen is probably made with the AdS/CFT correspondence in mind, as the boundary where the CFT lives is a timelike surface in AdS. Verlinde shows that, by associating a suitable entropy flux with the matter entering the screen and the Unruh temperature with the acceleration of the observers living on the screen, the Newton law -- and its relativistic extension to static spacetimes -- is implied by the first law of thermodynamics. 

The purpose of this note is to extend Jacobson's argument to timelike surfaces by calculating the lensing effect that a flux of incoming matter has on the accelerating observers living on a timelike screen.  A congruence of null geodesics is entirely specified by the initial conditions. 
Since timelike observers are much less constrained, we will have to be more pedantic about their behavior. We show that, if few geometrical conditions are met, Eq. \eqref{first} applies straightforwardly also to a screen made by timelike observers of finite acceleration. The change of internal energy $dE$ should be interpreted as the flux of energy momentum through the surface element $\Delta {\cal A}$ in the proper time $d s$, $T=A/2 \pi$ is the Unruh temperature of the observers of acceleration $A$ living on the screen and $dS$ is related with the infinitesimal change of area by the usual black hole formula $dS = d(\Delta {\cal A})/4 G$. We refer mainly to the General Relativity books~\cite{haw,poisson} for the calculations that follow. Also, we should mention the work~\cite{ash}, as it shares some similarities with the present approach.

\section{Building the screen}

The congruence of observers is defined by the timelike unit vector field $V_a$, $V_a V^a = -1$. In every point we can define $h^a_{\ b} = \delta^a_{\ b} + V^a V_b$, the projector onto the subspace orthogonal to $V$. We assume that the vector field $V$ has zero vorticity, which means that we can write its derivative as
\begin{equation} \label{1}
V_{a;b} = \theta_{ab} - {\vd}_a V_b,
\end{equation}
where the symmetric tensor $\theta_{ab} = h^d_{\ b} V_{a;d}$ is the expansion and the acceleration vector $\vd_a = V^b V_{a;b}$ is orthogonal to $V$, $\vd^a V_a = 0$. 
We now introduce the timelike screen by giving the first two conditions: \\

\begin{quote}
(\emph{i}) The three-dimensional timelike hypersurface ${\cal H}$ is spanned by the worldlines of a two-parameters family of observers within the congruence.\\

(\emph{ii}) At each point/event on the hypersurface $\cH$ the acceleration of the observers is orthogonal to $\cH$.\\
\end{quote} 
The last requirement implies that the unit normal to the surface $n_a$, $n_a n^a = 1$, is   proportional to $\vd_a$. By defining the acceleration scalar $A$ as $A^2 = \vd^a \vd_a$, we thus have 
\begin{equation} \label{proportional}
A \, n_a = \vd_a .
\end{equation}
With $n$ we can define another projector tensor, this time onto the surface ${\cal H}$: ${\bar h}^a_{\ b} = \delta^a_{\ b} - n^a n_b$.
It is tempting to try and ask that the acceleration be the same on the entire surface. We will see that asking that neighboring observers have the same acceleration is too strong a requirement which cannot be realized in general.  However, we can ask that the acceleration be preserved along each worldline. Since we are in the presence of non-geodesic observers, the appropriate quantity to be set to zero is the Fermi derivative~\cite{haw} of $\vd$:\\

\begin{quote}
(\emph{iii}) Along each timelike curve the acceleration vector is Fermi-transported:
\begin{equation} \label{fermi}
\frac{D_F \vd_a}{\partial s} = V^b \vd_{a;b} - A^2 V_a = 0 .
\end{equation}
\end{quote}
The absence of vorticity, eq.~\eqref{1}, guarantees~\cite{poisson} that $V$ is hypersurface orthogonal. This defines on ${\cal H}$ a set of ``equal-time" two-dimensional spacelike surfaces each of which we will denote with ${\cal S}$. The space tangent to ${\cal S}$ is reached by applying in turn the two projectors $h^a_{\ b}$ and ${\bar h}^a_{\ b}$. This defines a new projector,
\begin{equation}
q^a_{\ b} = h^a_{\ c} {\bar h}^c_{\ b} = m^a m_b + l^a l_b.
\end{equation}
In the last equality we have introduced other two unit vectors to complete the orthonormal basis. Of $n$, $V$, $m$ and $l$, the first is orhogonal to ${\cal H}$ and the first two are orthogonal to ${\cal S}$.

A key ingredient of our derivation is the extrinsic curvature of ${\cal H}$, $K_{ab} = {\bar h}^c_{\ a} {\bar h}^d_{\ b} n_{c;d}$. It is useful to express such a tensor in terms of the expansion properties of our observers. Since $n$ is related to $\vd$ by~\eqref{proportional} we first have to calculate the derivatives of the latter. A straightforward calculation gives:
\begin{eqnarray} 
\vd_{a;b} & = & V^c_{\ ;b} V_{a;c} + V^c V_{a;c b} \nonumber \\
&=& V^c_{\ ;b} V_{ a;c} + V^c V_{a;bc} + R_{acbd} V^c V^d \nonumber \\
&=& \theta^c_{\ b} \theta_{ac} - \theta_{ac} \vd^c V_b + V^c V_{a;bc} + R_{acbd} V^c V^d\, .\label{2}
\end{eqnarray}
By imposing \eqref{fermi} we obtain the relation 
\begin{equation}
\theta_{ac} \vd ^c = A^2 V_a -  V^b V^c V_{a;bc} ,
\end{equation}
that can be inserted back into \eqref{2}, giving
\begin{equation}\label{3}
\vd_{a;b} = - A^2 V_a V_b + \theta^c_{\ b} \theta_{ac} + V^c h^d_{\ b} V_{a;dc} + S_{ab}\, ,
\end{equation}
where  we have defined the symmetric tensor $S_{ab} = R_{acbd} V^c V^d$. By the cyclic property of the Riemann tensor, $S_{ab}$ has components only in the subspace orthogonal to $V$.
The third term on the RHS of \eqref{3} can be further decomposed:
\begin{equation} \label{5}
V^c h^d_{\ b} V_{a;dc} = \td_{ab} - \vd_a \vd_b - \theta_{ac} \vd^c V_b\, ,
\end{equation}
where $\td_{ab} \equiv V^c (V_{a;d} h^d_{\ b})_{;c} = V^c \theta_{ab;c}$ and the projector tensor $h^a_{\ b}$ is meant to be Fermi-transported along the curve. 

In order for condition (\emph{ii}) to be consistent
we now should impose that the vector field $\vd_a$ be hypersurface orthogonal. By Frobenius' theorem this is equivalent to saying that 
\begin{equation} \label{4}
\vd_{[a;b} \vd_{c]} =0 .
\end{equation}
The symmetric terms of $\vd_{a;b}$ automatically satisfy \eqref{4}, so the only constraint comes from imposing \eqref{4} on the asymmetric term in \eqref{5}:
\begin{equation}
\vd^c \theta _{c[a} V_b \vd_{c ]} = 0,
\end{equation}
from which we argue that the vector $\vd^c \theta _{ca}$ has to be parallel to $\vd_a$. In other words, the expansion tensor can be decomposed as 
\begin{equation}
\theta_{ab} = {\tilde \theta}_{ab} + \theta_3 n_a n_b\, ,
\end{equation}
where the crucial quantity ${\tilde \theta}_{ab}$, defined on the tangent space of ${\cal S}$, is the two-dimensional expansion of the observers living on the screen ${\cal H}$. More generally, we will always indicate with a ``tilde" quantities projected on the tangent space of ${\cal S}$, say, ${\tilde X}_{ab} = q^c_{\ a} q^d_{\ b} X_{cd}$. The expansion along the $n$ direction, $\theta_3$, will be irrelevant for the argument of this paper.

The variation of the acceleration $A$ along ${\cal H}$ is ultimately enforced by condition 
(\emph{ii}) and cannot be decided a priory. In order to see this, we differentiate the identity  
$n_a n^a =1$ and project it along the surface, $(n_a n^a)_{;b}{\bar h}^b_{\ c} = 0$. We then express $n$ by using~\eqref{proportional}. Along the $V$ direction such a condition is satisfied by construction, as the acceleration has been Fermi transported along the curves. Along ${\cal S}$ is a different story, the condition gives
\begin{equation}
\frac{1}{A}(n^c \vd_{c;b} - A_{;b})q^b_{\ a} = 0.
\end{equation}
The first term inside the brackets picks up the off diagonal terms ($n-m$ or $n-l$) of $\vd_{a;b}$. By looking at equation \eqref{3} such terms are entirely determined by $S_{ab}$, whose value depends on the Riemann tensor. In summary, the behavior of the function $A$ along ${\cal S}$ is fixed by the off diagonal terms of $S_{ab}$: 
\begin{equation}
 A_{;a} = n^c S_{cb} q^b_{\ a} .
\end{equation}

\section{The Gauss-Codazzi equation on the timelike screen}

By normalizing \eqref{3} and projecting onto the surface we can finally write the extrinsic curvature of ${\cal H}$,
\begin{equation} \label{extrinsic}
K_{ab} = \frac{1}{A} \left(- A^2 V_a V_b + {\dot{\tilde \theta}}_{ab} + {\tilde \theta}^c_{\ b} {\tilde \theta}_{ac} + {\tilde S}_{ab} \right) \equiv - \, A\,  V_a V_b + {\tilde K}_{ab},
\end{equation}
where a dot, here and in the following, means derivative in the $V$ direction.
The Fermi derivative applies to the vectors $m$ and $l$  just as a normal derivative, because, by construction, they are orthogonal to both the velocity $V$ and the acceleration $\vd$. Therefore, $m$ and $l$ are parallely transported along the curves. As a result, time derivatives commute with projections on ${\cal S}$ and in the above equation we have used  ${\tilde \td}_{ab} = {\dot{\tilde \theta}}_{ab}$.
By looking at equation \eqref{extrinsic}, we note that the projected extrinsic curvature of ${\cal H}$  can be chosen arbitrarily at any given time by specifying the derivative of the expansion, ${\dot{\tilde \theta}}_{ab}$. The derivatives of $K_{ab}$, instead, are constrained by the Gauss Codazzi equation~\cite{poisson},
\begin{equation} \label{GC}
 (R_{dabc}n^d)_{\parallel {\cal H}} = (K_{ab;c} - K_{ac;b})_{\parallel {\cal H}},
\end{equation}
where the subscript ``${\parallel {\cal H}}$" means that the tensor is projected along ${\cal H}$.
The above relation allows to relate the geometrical property of the surface with the matter flux through it. 

The flux of matter energy through the area element $\Delta {\cal A}$ of the surface, during the (proper) time interval $d s$ reads
$d E = T_{ab} V^a n^a \, \Delta{\cal A} \, ds.$
By Einstein's equations this is just\footnote{As noted already in~\cite{jac} and ~\cite{pad}, curiously enough, when using such thermodynamics arguments, the cosmological constant term in the Einstein equations drops.}
\begin{equation}
d E = \frac{1}{8 \pi G}\, R_{ab} V^a n^a \, \Delta{\cal A} \, ds.
\end{equation}
The scalar quantity in the last equation can be calculated with the Gauss-Codazzi equation~\eqref{GC}, 
\begin{equation} \label{basic}
R_{ab} V^a n^a  = q^{ab} V^c (K_{ac;b}-K_{ab;c}) .
\end{equation}
The first term in the brackets is obtained by use of condition (\emph{iii}), eq.~\eqref{fermi},
\begin{eqnarray}
q^{ab} V^c K_{ac;b} &=& q^{ab} [(V^c K_{ac})_{;b} - V^c_{\ ;b} K_{ac}] \nonumber \\
&=& q^{ab} [(A V_a)_{;b} - (\theta^c_{\ b} - \vd^c V_b) K_{ac}]\nonumber \\
&=& q^{ab} [A_{;b}V_a + A (\theta_{ab} - \vd_a V_b) - \theta^c_{\ b} - \vd^c V_b) K_{ac}] \nonumber \\
&=& A \, {\tilde \theta} - {\tilde \theta}^c_{\ b} {\tilde K}_{ac}\, ,
\end{eqnarray}
where ${\tilde \theta} \equiv q^{ab} {\tilde \theta}_{ab}$ is the trace of the expansion along ${\cal H}$ and can be interpreted as the log derivative of the area element spanned by the observers along ${\cal H}$ with respect to their proper time: ${\tilde \theta} = d \ln (\Delta {\cal A})/d s$.
The second term in the brackets of the RHS of \eqref{basic} is the trace of the derivative of the extrinsic curvature along $V$, 
\begin{equation} \label{kdot}
{\dot K}_{ab} = -A(\vd_a V_b + V_a \vd_b) + \frac{1}{A}({\ddot {\tilde \theta}}_{ab} + {\dot {\tilde \theta}^d}_{b}{\tilde \theta}_{ad} + {\tilde \theta^d}_{\ b} {\dot {\tilde \theta}}_{ad} + {\dot {\tilde S}}_{ab}),
\end{equation}
projected onto ${\cal S}$. 

\subsection{The flat screen}

Something interesting happens if ${\tilde K}_{ab} = 0$ everywhere on the surface. In this case, $q^{ab} V^c K_{ac;b} =  A \, {\tilde \theta}$, $q^{ab} V^c K_{ab;c} = 0$ by construction and thus $R_{ab}V^a n^b =  A \, {\tilde \theta}$. In other words, 
\begin{equation} \label{result}
d E = \frac{1}{8 \pi G} A \, {\tilde \theta} \, \Delta {\cal A}\, ds = \frac{1}{8 \pi G} A \, d (\Delta {\cal A)}.
\end{equation}
The latter is equivalent to~\eqref{first} upon the usual identification $T = A/2\pi$ (the Unruh temperature of the observer with acceleration $A$) and $S = {\cal A}/4 G$ (the usual entropy area law). The last condition that we need to impose therefore reads\\

\begin{quote}
(\emph{iv}-a) The projection of the extrinsic curvature is everywhere null on ${\cal H}$:
\begin{equation}
{\tilde K}_{ab} \equiv {\dot{\tilde \theta}}_{ab}  + {\tilde \theta^c}_{\ b} {\tilde \theta}_{ac} + {\tilde S}_{ab} =0\, .
\end{equation}\\
\end{quote}

\subsection{The ``cone"}

Alternatively, and more generally, we can obtain eq.~\eqref{result} by allowing an arbitrary ``initial" projected extrinsic curvature ${\tilde K}_{ab}$ and imposing that the surface evolves along the time flow in such a way that
\begin{equation} \label{cone}
\frac{D_F {\tilde K}_{ab}}{\partial s} \equiv V^c {\tilde K}_{ab;c} = - {\tilde \theta}^c_{\ b} {\tilde K}_{ac}.
\end{equation}
By using~\eqref{kdot}, it is straightforward to translate the above relation into a condition for ${\ddot {\tilde \theta}}$. However, it is perhaps more interesting to look at its geometrical meaning. 
In the case of zero expansion, ${\tilde \theta}_{ab}=0$, the RHS term vanishes and the above condition is just saying that ${\cal H}$ is, locally, a \emph{cylinder} from the point of view of its extrinsic properties: it has an initial extrinsic curvature  ${\tilde K}_{ab}$ and such a quantity is preserved during the time evolution. In the case of non zero expansion, instead, eq.~\eqref{cone} suggests that the extrinsic curvature properties are rescaled in time proportionally to the expansion of the observers.
As the following back on the envelope argument shows, from the point of view of the extrinsic curvature, we are moving, locally, along a \emph{cone}.

Consider a two dimensional cone in three flat dimensions. Any distance $s$ from the vertex defines a one-dimensional section of the cone: a circle of radius $R(s)$. The extrinsic curvature along the cone is therefore ${\tilde k}(s) = 1/R(s)$. Thus we have 
\begin{equation}\label{cone2}
\frac{d{\tilde k}}{d s} =  -\frac{1}{R^2} R'(s) = - \Theta \, {\tilde k} ,
\end{equation}
where $\Theta = R'(s)/R(s)$ is just the one-dimensional expansion of the straight lines passing through the vertex. Eq. \eqref{cone} is nothing but a higher dimensional generalization of~\eqref{cone2}. The more general condition thus reads\\

\begin{quote}
(\emph{iv}-b) The extrinsic properties of ${\cal H}$ are, locally, those of a three-dimensional cone of expansion ${\tilde \theta}_{ab}$. The time behavior of the extrinsic curvature ${\tilde K}_{ab}$ is related to the expansion by  
\begin{equation}
{\dot {\tilde K}}_{ab} = - {\tilde \theta}^c_{\ b} {\tilde K}_{ac}.
\end{equation}\\
\end{quote}

\section{Conclusions}

We have enquired under which conditions the result of Jacobson~\cite{jac} can be extended to a timelike screen of observers of finite acceleration. Jacobson's construction only requires that the initial expansion of the null generators be zero (``equilibrium condition"). By applying the Raychaudhuri equation, he finds that the derivative of such an expansion is bound to follow the first law of thermodynamics~\eqref{first}, once an appropriate dictionary is used to translate between geometrical and thermodynamical quantities. Our construction is inevitably more involved -- as timelike observers have more freedom than null geodesics -- but  interesting because it relates directly the expansion of the observers (not its derivative!) with the flux of matter through the screen. Hence, we can follow the ``thermodynamics" of that bunch of observers indefinitely, whereas Jacobson's null generators cease to be useful as soon as they acquire a non-zero expansion. Our results, involving the expansion directly, have the character of a constraint, rather than that of a dynamical property. The expansion is positive if the flux of matter is in the same direction of the acceleration and negative otherwise. 

A sensible thermodynamic result is found by finally imposing either of two ``regularity" conditions on the screen. Condition (\emph{iv}-b) is obviously more general as it contains 
(\emph{iv}-a) in the appropriate limit. Condition (\emph{iv}-a), however, is particularly neat as it does not rely on the knowledge of the expansion of the observers inside the surface, but only on the extrinsic properties of the latter. 

Our results apply to general spacetimes. However, in order to get some intuition, it is helpful to consider a stationary self gravitating system (such as a star) and build a ``screen" of stationary observers all around it~\cite{ver}. Such observers have precisely the right amount of acceleration that allow them to stay at a given distance from the object. If we perturb the system by throwing in some matter, the total mass inside the screen increases. In this work we found how the screen locally back-reacts to the incoming matter, under the condition that the observers keep the same acceleration. As long as no matter falls in, the observers worldlines will continue to describe a ``cylinder", the extrinsic curvature of the screen being conserved in time. In the presence of a flux of matter coming from infinity, instead, they start converging and will eventually fall towards the center of the gravitating system. We have shown that such a ``converging" effect is completely local and consistently contained in the Einstein equations. On the opposite, if matter is thrown out of the screen (``out" here means, more generally,  in the same direction as the acceleration) the observers expand and tend to escape from the gravitational field of the object.

The analogies between gravity and thermodynamics are sometimes taken as hints that gravity is an emergent phenomenon, possibly together with spacetime itself~\cite{glimmers,olaf,review}. Possible deeper interpretations of the simple General Relativity result
that we have found here are left for future work. \\

{\bf Acknowledgments}
I am indebted to Olaf Dryer for  many exciting discussions and his collaboration during the early stages of this work. I also thank Badri Krishnan for making me aware of Ref.~\cite{ash}. The research at Perimeter Institute is supported in part by the Government of Canada through NSERC and by the Province of Ontario through
the Ministry of Research \& Innovation.\\


\begin{thebibliography}{99}
 
 \bibitem{bek}
  J.~D.~Bekenstein,
  ``Black holes and entropy,''
  Phys.\ Rev.\  D {\bf 7}, 2333 (1973).
 
 \bibitem{bardeen}
 J.~M.~Bardeen, B.~Carter and S.~W.~Hawking,
  ``The Four laws of black hole mechanics,''
  Commun.\ Math.\ Phys.\  {\bf 31}, 161 (1973).
 
 \bibitem{jac} 
  T.~Jacobson,
  ``Thermodynamics of space-time: The Einstein equation of state,''
  Phys.\ Rev.\ Lett.\  {\bf 75}, 1260 (1995)
  [arXiv:gr-qc/9504004].
 
 \bibitem{lev}
  A.~V.~Frolov and L.~Kofman,
  ``Inflation and de Sitter thermodynamics,''
  JCAP {\bf 0305}, 009 (2003)
  [arXiv:hep-th/0212327].
  
  \bibitem{cai}
  R.~G.~Cai and S.~P.~Kim,
  ``First law of thermodynamics and Friedmann equations of
  Friedmann-Robertson-Walker universe,''
  JHEP {\bf 0502}, 050 (2005)
  [arXiv:hep-th/0501055].

 
 \bibitem{pad}
  T.~Padmanabhan,
  ``Gravity and the thermodynamics of horizons,''
  Phys.\ Rept.\  {\bf 406}, 49 (2005)
  [arXiv:gr-qc/0311036];
  T.~Padmanabhan,
  ``Thermodynamical Aspects of Gravity: New insights,''
  Rept.\ Prog.\ Phys.\  {\bf 73}, 046901 (2010)
  [arXiv:0911.5004 [gr-qc]];
    S.~Kolekar and T.~Padmanabhan,
  ``Holography in Action,''
  arXiv:1005.0619 [gr-qc].
  
  \bibitem{ver}
  E.~P.~Verlinde,
  ``On the Origin of Gravity and the Laws of Newton,''
  arXiv:1001.0785 [hep-th].
  
  \bibitem{nic}
  P.~Nicolini,
  ``Entropic force, noncommutative gravity and ungravity,''
  Phys.\ Rev.\  D {\bf 82}, 044030 (2010)
  [arXiv:1005.2996 [gr-qc]].
  
    
   \bibitem{glimmers}
  F.~Piazza,
  ``Glimmers of a pre-geometric perspective,''
  Found.\ Phys.\  {\bf 40}, 239 (2010)
  [arXiv:hep-th/0506124].

  \bibitem{sergio}
  S.~Cacciatori, F.~Costa and F.~Piazza,
  ``Renormalized Thermal Entropy in Field Theory,''
  Phys.\ Rev.\  D {\bf 79}, 025006 (2009)
  [arXiv:0803.4087 [hep-th]].
  
   \bibitem{review}
  F.~Piazza,
  ``New views on the low-energy side of gravity,''
  arXiv:0910.4677 [gr-qc].

    \bibitem{haw}
  S.~W.~Hawking and G.~F.~R.~Ellis,
  ``The Large scale structure of space-time,''
{\it  Cambridge University Press, Cambridge, 1973}

\bibitem{poisson}
E.~Poisson,
``A Relativist's Toolkit: The Mathematics of Black-Hole Mechanics,"
 {\it  Cambridge University Press, Cambridge, 1973}
 
 \bibitem{ash}
 A.~Ashtekar and B.~Krishnan,
  ``Dynamical horizons and their properties,''
  Phys.\ Rev.\  D {\bf 68}, 104030 (2003)
  [arXiv:gr-qc/0308033].
 
    
  \bibitem{olaf}
  O.~Dreyer,
  ``Emergent general relativity,''
  arXiv:gr-qc/0604075.
  
 
 
\end{thebibliography}
\end{document}